\documentclass{article}
\usepackage[preprint,nonatbib]{neurips_2020}

\usepackage{neurips_2020}
\usepackage{acronym}
\usepackage{amsmath}
\usepackage{graphicx}
\usepackage[dvipsnames]{xcolor}

\usepackage[utf8]{inputenc} 
\usepackage[T1]{fontenc}    
\usepackage{url}            
\usepackage{booktabs}       
\usepackage{amsfonts}       
\usepackage{nicefrac}       
\usepackage{microtype}      
\usepackage{subcaption}

\usepackage{physics}
\usepackage{amsmath}
\usepackage{tikz}
\usepackage{mathdots}
\usepackage{yhmath}
\usepackage{cancel}
\usepackage{color}
\usepackage{siunitx}
\usepackage{array}
\usepackage{multirow}
\usepackage{amssymb}
\usepackage{gensymb}
\usepackage{tabularx}
\usepackage{booktabs}
\usetikzlibrary{fadings}
\usetikzlibrary{patterns}
\usetikzlibrary{shadows.blur}
\usetikzlibrary{shapes}

\newcommand{\argmax}{\operatornamewithlimits{argmax}}

\newcommand{\comment}[1]{}

\acrodef{TF-DOAnet}{time-frequency direction-of-arrival net}
\acrodef{WDO}{W-disjoint orthogonality}
\acrodef{SOTA}{state-of-the-art}
\acrodef{FCN}{fully convolutional network}
\acrodef{DDESS}{deep direction estimation for speech separation }
\acrodef{NMF}{non-negative matrix factorization}
\acrodef{SIR}{signal to noise ratio}
\acrodef{SDR}{signal to distortion ratio}
\acrodef{TF}{time-frequency}
\acrodef{BF}{beamformer}
\acrodef{BSS}{blind source separation}
\acrodef{DOA}{direction of arrival}
\acrodef{SPI}{speaker position identifier}
\acrodef{GSC}{general sidelobe canceller}
\acrodef{DSE}{Deep Single Expert}
\acrodef{MVDR}{ minimum variance distortionless response}
\acrodef{GEVD}{generalized eigenvalue decomposition}
\acrodef{AIR}{acoustic impulse response}
\acrodef{PSD}{power spectral density}
\acrodef{cPSD}{cross-power spectral density}
\acrodef{FIR}{finite-impulse response}
\acrodef{MTF}{multiplicative transfer function}
\acrodef{RIR}{room impulse response}
\acrodef{LTI}{linear time invariant}
\acrodef{DNN}{deep neural network}
\acrodef{CNN}{convolutional neural network}
\acrodef{MFCC}{mel-frequency cepstral coefficients}
\acrodef{MMSE}{minimum mean square error}
\acrodef{ASR}{automatic speech recognition}
\acrodef{ATF}{acoustic transfer function}
\acrodef{LCMV}{linearly constrained minimum variance}
\acrodef{RTF}{relative transfer function}
\acrodef{iRTF}{instantaneous relative transfer function}
\acrodef{VAD}{voice activity detector}
\acrodef{STSA}{short-time spectral amplitude estimator}
\acrodef{LSAE}{log-spectral amplitude estimator}
\acrodef{OMLSA}{optimally modified log spectral amplitude}
\acrodef{IMCRA}{improved minima controlled recursive averaging}
\acrodef{STFT}{short-time Fourier transform}
\acrodef{DFT}{discrete Fourier transform}
\acrodef{MoG}{Mixture of Gaussians}
\acrodef{MOE}{Mixture of Experts}
\acrodef{MODE}{Mixture of Deep Experts}
\acrodef{r.v.}{random variable}
\acrodef{p.d.f.}{probability density function}
\acrodef{NN}{neural network}
\acrodef{EM}{expectation-maximization}
\acrodef{SPP}{speech presence probability}
\acrodef{CMVN}{cepstral mean and variance normalization}
\acrodef{NN-MM}{neural network mixture-maximum}
\acrodef{PESQ}{perceptual evaluation of speech quality}
\acrodef{SNR}{signal to noise ratio}
\acrodef{SIR}{signal to interference ratio}
\acrodef{DAE}{deep auto-encoder}
\acrodef{LLR}{log likelihood ratio}
\acrodef{WSS}{weighted spectral slope}
\acrodef{Covl}{overall quality}
\acrodef{Csig}{speech distortion}
\acrodef{Cbak}{background distortion}
\acrodef{WSJ}{Wall Street Journal}
\acrodef{WSJ1}{Wall Street Journal 1}

\acrodef{SVM}{support vector machine}
\acrodef{IBM}{ideal binary mask}
\acrodef{IRM}{ideal ratio mask}
\acrodef{ReLU}{rectified linear unit}
\acrodef{WER}{word error rate}
\acrodef{MM}{MixMax}
\acrodef{MOS}{mean opinion score}
\acrodef{mse}{mean squre error}
\acrodef{pDNN}{phoneme DNN}
\acrodef{cDNN}{classifier DNN}
\acrodef{SGD}{stochastic gradient descent}
\acrodef{TDOA}{time difference of arrival}
\acrodef{CSD}{concurrent speakers detector}
\acrodef{STOI}{short-time objective intelligibility measure}
\acrodef{MCCSD}{multichannel concurrent speakers detector}
\acrodef{RIR}{room impulse response}
\acrodef{WSJ}{wall street journal}
\acrodef{BLSTM}{bidirectional long short-term memory}
\acrodef{GCC}{generalized cross correlation}
\acrodef{SRP-PHAT}{steered response power with phase transform}
\acrodef{BIU}{Bar-Ilan University}
\acrodef{ULA}{uniform linear array}
\acrodef{MAE}{mean absolute error}
\acrodef{MUSIC}{multiple signal classification}
\acrodef{w.r.t.}{with respect to}
\acrodef{CMS-DOA}{CNN multi-speaker DOA}

\title{FCN Approach for Dynamically Locating Multiple Speakers}

\author{Hodaya Hammer\\
  Department of Electrical Engineering\\
  Bar-Ilan University\\
  Ramat-Gan, 5290002\\
  Israel\\
  \texttt{hodib91@gmail.com} \\
  \And Shlomo E. Chazan \\
  Department of Electrical Engineering\\
  Bar-Ilan University\\
  Ramat-Gan, 5290002\\
  Israel\\
  \texttt{Shlomi.Chazan@biu.ac.il} \\
  \And
  Jacob Goldberger \\
  Department of Electrical Engineering\\
  Bar-Ilan University\\
  Ramat-Gan, 5290002\\
  Israel\\
  \texttt{jacob.goldberger@biu.ac.il} \\
  \And
  Sharon Gannot \\
  Department of Electrical Engineering\\
  Bar-Ilan University\\
  Ramat-Gan, 5290002\\
  Israel\\
  \texttt{Sharon.Gannot@biu.ac.il} \\
}

\begin{document}

\maketitle

\begin{abstract}
In this paper, we present a deep neural network-based online multi-speaker localisation algorithm. Following the W-disjoint orthogonality principle in the spectral domain, each \ac{TF} bin is dominated by a single speaker, and hence by a single \ac{DOA}. A fully convolutional network is trained with instantaneous spatial features to estimate the \ac{DOA} for each \ac{TF} bin. The high resolution classification enables the network to accurately and simultaneously localize and track multiple speakers, both static and dynamic. Elaborated experimental study using both simulated and real-life recordings in static and dynamic scenarios, confirms that the proposed algorithm outperforms both classic and recent deep-learning-based algorithms.
\end{abstract}

\section{Introduction}
    
Locating multiple sound sources recorded with a microphone array in an acoustic environment is an essential component in various cases such as source separation and  scene analysis.  The relative location of a sound source with respect to a microphone array is generally given in the term of the \ac{DOA} of the sound wave originating from that location.  \ac{DOA} estimation and tracking are generating interest lately, due to the need for far-field enhancement and recognition in smart home devices.
In real-life environments, sound sources are captured by the microphones together with acoustic reverberation. While propagating in an acoustic enclosure, the sound wave undergoes reflections from the room facets and from various objects. These reflections deteriorate speech quality and, in extreme cases, its intelligibility. Furthermore, reverberation increases the time dependency between speech frames, making source \ac{DOA} estimation a very challenging task.

A plethora of classic signal processing-based approaches have been proposed throughout the years for the task of broadband \ac{DOA} estimation.  
The \ac{MUSIC} algorithm~\cite{schmidt_multiple_1986} applies a subspace method that was later adapted to the challenges of speech processing in \cite{Dmochowski_Benesty_Affes_2007}.
The \ac{SRP-PHAT} algorithm \cite{Brandstein_Silverman_1997} uses generalizations of  cross-correlation methods for \ac{DOA} estimation. 
These methods are still widely in use. However, in high reverberation enclosures, their performance is not satisfactory. 


Supervised learning methods encompass an advantage for this task since they are data-driven. Deep-learning methods can be trained to find the \ac{DOA} in different acoustic conditions. Moreover, if a network is trained using rooms with different acoustic conditions and multiple noise types, it can be made robust against noise and reverberation even for rooms which were not in the training set.
Deep learning methods have recently been proposed for sound source localization.
In \cite{Xiao_Zhao_Zhong_Jones_Chng_Li_2015,Vesperini_Vecchiotti_Principi_Squartini_Piazza_2016} simple feed-forward \acp{DNN} were trained using \ac{GCC}-based audio features, demonstrating improved performance as compared with classic approaches. Yet, this method is mainly designed to deal with  a single sound source at a time. In 
\cite{Takeda_Komatani_2016} the authors trained a DNN for multi-speaker \ac{DOA} estimation. In high reverberation conditions, however, their performance is not satisfactory. In~\cite{Pujol_Bavu_Garcia_IRNIA_2019,vera2018towards}  time domain features were  used and they have shown performance improvement in highly-reverberant enclosures. 
In~\cite{sumitru2017_doaSingle}, a \ac{CNN} based classification method was applied in the \ac{STFT} domain for broadband \ac{DOA} estimation, assuming that only a single speaker is active per time frame. The phase component of the \ac{STFT} coefficients of the input signal were directly provided as input to the \ac{CNN}. This work was extended in~\cite{sumitru2019_doaMulti} to estimate multiple speakers' \acp{DOA}, and has shown high \ac{DOA} classification performances. In this approach, the \ac{DOA} is estimated for each frame independently.
The main drawback of most DNN-based approaches, however, is that they only use low-resolution supervision, namely only time frame or even utterance-based labels. In speech signals, however, each time-frequency bin is dominated by a single speaker, a property referred to as \ac{WDO} \cite{w_disjoint}.  Adopting this model results in higher resolution, which might be beneficial for the task at hand.
This model was also utilized in~\cite{chazan_hammer_hazan_goldberger_gannot_2019} for speech separation  where the authors recast the separation problem as a \ac{DOA} classification at the \ac{TF} domain. A \ac{FCN} was trained using spatial features to infer the \ac{DOA} at every \ac{TF} bin. Although the \ac{DOA} resolution was relatively low, it was sufficient for the separation task at low reverberation conditions. 
When applying this method in high-reverberation enclosures or to separate adjacent speakers, a performance degradation was observed.

    
In this work, we present a multi-speaker \ac{DOA} estimation algorithm. According to the \ac{WDO} property of speech signals \cite{w_disjoint,yilmaz2004blind}, each \ac{TF} bin is dominated by (at most) a single speaker. This \ac{TF} bin can therefore be associated with a single \ac{DOA}. 
We use instantaneous spatial cues from the microphone signals. These features are used to train a \ac{FCN} to infer the \ac{DOA} of each \ac{TF} bin. The \ac{FCN} is trained to address  various reverberation conditions. The \ac{TF}-based classification facilitates the tracking ability for multiple moving speakers. In addition, unlike many other supervised domains, the \ac{DOA} domain lacks a standard benchmark. The LOCATA dataset \cite{locata} was recorded in one room with relatively low reverberation (RT$_{60}=0.55$). Furthermore, a training dataset with high \ac{TF} labels is not publicly available.  Therefore,  we generated training and test datasets simulating various real-life scenarios. 
We tested the proposed method on simulated data, using publicly available \acp{RIR} recorded in a real room~\cite{Elior_database}, as well as real-life experiments. 
We show that the proposed algorithm significantly outperforms state-of-the-art competing methods.



The main contribution of this paper is the A high resolution \ac{TF}-based approach that improves  \ac{DOA} estimation performances \ac{w.r.t.}~the \ac{SOTA} approaches, which are frame-based, and enables simultaneously tracking multiple moving speakers.

\section{Multiple speaker' location algorithm}
\label{sec:Problem_formulation}

\subsection{Time-frequency features}
\label{subsec:feature_vector}
Consider an array with $M$ microphones acquiring a mixture of $N$ speech sources in a reverberant environment. The $i$-th speech signal  $s^i(t)$ propagates through the acoustic channel before being acquired by the $m$-th microphone:
\begin{equation}
z_m(t)=\sum_{i=1}^{N} s^i(t)*h_m^i(t), \hspace{0.8cm} m =1,\ldots,M
\label{problem_time},
\end{equation}
where $h_m^i$ is the \ac{RIR} relating the $i$-th speaker and the $m$-th microphone. In the \ac{STFT} domain \eqref{problem_time} can be written as (provided that the frame-length is sufficiently large \ac{w.r.t.}~the filter length):
\begin{equation}
z_m(l,k)=\sum_{i=1}^{N} s^i(l,k)h_m^i(l,k),
\label{eq:problem}
\end{equation}
where $l$ and $k$, are the time frame and the frequency indices,  respectively.

The \ac{STFT} \eqref{eq:problem} is complex-valued and hence comprises both spectral and phase information. It is clear that the spectral information alone is insufficient for \ac{DOA} estimation. It is therefore a common practice to use the phase of the \ac{TF} representation of the received microphone signals, or their respective phase-difference, as they are directly related to the \ac{DOA} in non-reveberant environments.

We decided to use an alternative feature, which is generally independent of the speech signal and is mainly determined by the spatial information. For that, we have selected the \ac{RTF} \cite{gannot2001RTF} as our feature, since it is known to encapsulate the spatial fingerprint for each sound source. Specifically, we use the \acf{iRTF}, which is the bin-wise ratio between the $m$-th microphone signal and the reference microphone signal $z_\text{ref}(l,k)$:
\begin{equation}
    \label{rtf_formulation}
    \text{iRTF}(m,l,k) = \frac{z_m(l,k)}{z_\text{ref}(l,k)}.
\end{equation}
Note, that the reference microphone is arbitrarily chosen. Reference microphone selection is beyond the scope of this paper (see \cite{RTF_ref_mic_selec} for a reference microphone selection method). The input feature set extracted from the recorded signal is thus a 3D tensor $\mathcal{R}$: 
\begin{equation}
\mathcal{R}(l,k,m) = [\mathfrak{Re}(\text{iRTF}(m,l,k)), 
            \mathfrak{Im}(\text{iRTF}(m,l,k))
            ] .
  \label{eq:est_RTF}
\end{equation}
The matrix $\mathcal{R}$  is  constructed from $L \times K $ bins, where $L$ is the number of time frames and $K$ is the number of frequencies. Since the \acp{iRTF} are normalized by the reference microphone, it is excluded from the features. Then for each \ac{TF} bin $(l,k)$, there are  $P=2(M-1)$ channels, where the multiplication by $2$ is due to the real and imaginary parts of the complex-valued feature.
 For each \ac{TF} bin the spatial features were normalized to have a zero mean and a unit variance.

Recall that the \ac{WDO} assumption \cite{w_disjoint} implies that each \ac{TF} bin $(l,k)$ is dominated by a single speaker. 
Consequently, as the speakers are spatially separated, i.e. located at different \acp{DOA}, each \ac{TF} bin is dominated by a single \ac{DOA}. 

 Our goal in this work is to accurately estimate the speaker direction at every \ac{TF} bin from the given mixed recorded signal. 

\subsection{FCN for DOA estimation} 
We formulated the \ac{DOA} estimation  as a classification task by discretizing the \ac{DOA} range. The resolution was set to $5^\circ$, such that the \ac{DOA} candidates are in the set $\Theta=\{0^\circ,5^\circ,10^\circ,\ldots,180^\circ\}$.

Let $D_{l,k}$ be a \ac{r.v.} representing the active dominant direction, recorded at bin $(l,k)$. Our task boils down to deducing the conditional distribution of the discrete set of \acp{DOA} in $\Theta$ for each \ac{TF} bin, given the recorded mixed signal:
\begin{equation}
    \label{eq:theta_prob_r}
        p_{l,k}(\theta) = p( D_{l,k}=\theta|{\mathcal{R}}), \quad  \theta \in \Theta. 
\end{equation}
For this task, we use a \ac{DNN}. 
The network output is an $L\times K\times |\Theta|$ tensor, where $|\Theta|$ is the cardinality of the set $\Theta$.
Under this construction of the feature tensor and output probability tensor, a pixel-to-pixel approach for mapping a 3D  input `image', $\mathcal{R}$ and a 3D output `image', $ p_{l,k}(\theta)$, can be utilized.  
 An \ac{FCN} is used to compute \eqref{eq:theta_prob_r} for each \ac{TF} bin. The pixel-to-pixel method is beneficial in two ways. First, for each TF bin in our input image the network estimates the \ac{DOA} distribution separately. Second, the \ac{TF} supervision is carried out with the spectrum of the different speakers. The \ac{FCN} hence takes advantage of the spectral structure and the continuity of the sound sources in both the time and frequency axes. These structures contribute to the pixel-wise classification task, and prevent discontinuity in the \ac{DOA} decisions over time.  
In our implementation, we used a U-net architecture, similar to the one described in~\cite{unet_segmentation}. We dub our algorithm \acf{TF-DOAnet}.

The input to the network is the feature matrix $\mathcal{R}$ (\ref{eq:est_RTF}). 
In our U-net architecture, the input  shape is $(L,K,P)$ where $K=256$ is the number of frequency bins, $L=256$ is the number of frames, and $P=2M-2$ where $M$ is the number of microphones. The overlap between successive \ac{STFT} frames is set to $75\%$. This allows to improve the estimation accuracy of the \acp{RTF}, by averaging  three consecutive frames both in the numerator and denominator of \eqref{rtf_formulation}, without  sacrificing the instantaneous nature of the RTF.

\ac{TF} bins in which there is no active speech are non-informative. Therefore, the estimation is carried out only on speech-active  \ac{TF} bins. As we assume that the acquired signals are noiseless, we define a \ac{TF}-based \ac{VAD} as follows:
\begin{equation}
    \label{eq:vad}
        \text{VAD}(l,k) = \left\{
        \begin{array}{ll}
              1 & |z_\text{ref}(l,k)|\geq {\epsilon} \\
              0 & \text{o.w.} \\
        \end{array}
        \right.  ,
\end{equation}
where $\epsilon$ is a threshold value. In noisy scenarios, we can use a robust \ac{SPP} estimator instead of the \ac{VAD} \cite{wang2018supervised}.

\begin{figure*}[t!] 
	\centering
	\includegraphics[angle=-90,trim=130 120 300 0 ,clip,scale=0.6]{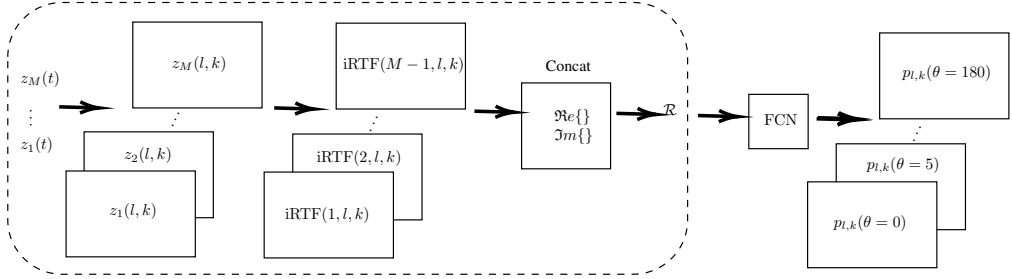}
	\caption{Block diagram of the \ac{TF-DOAnet} algorithm. The dashed envelope describes the feature extraction step.}
	\label{fig:unet}
\end{figure*}

The task of \ac{DOA} estimation only requires time frame estimates. Hence, we aggregate over all active frequencies at a given time frame to obtain a frame-wise probability:
\begin{equation}
    p_l(\theta)=\frac{1}{K'}\sum_{k=1}^K p_{l,k}(\theta)\text{VAD}(l,k).
\end{equation}
where $K'$ is the number of active frequency bands at the $l$-th time frame.
We thus obtain for each time frame a posterior distribution over all possible \acp{DOA}. 
If the number of speakers is known in advance, we can choose the directions corresponding to the highest posterior probabilities. If an estimate of the number of speakers is also required, it can be determined by applying a suitable threshold. Figure \ref{fig:unet} summarizes the \ac{TF-DOAnet} in a block diagram.

\subsection{Training phase}
\label{subsec:training_phase}
The supervision in the training phase is based on the \ac{WDO} assumption in which  each TF bin is dominated by (at most) a single speaker.
The training is based on simulated data generated by a publicly availble \ac{RIR} generator software\footnote{Available online at \texttt{github.com/ehabets/RIR-Generator}}, efficiently implementing the image method~\cite{allen_image_1979}. A four microphone linear array was simulated with $(8,8,8)$~cm inter-microphones distances. Similar microphone inter-distances were used in the test phase. For each training sample, the acoustic conditions were randomly drawn from one of the simulated rooms of different sizes and different reverberation levels $\textrm{RT}_{60}$ as described in Table \ref{table:training_data_config}. The microphone array was randomly placed in the room in one out of six arbitrary positions.

\begin{table}[t!]
    \centering
    \caption{Configuration of training data generation. All rooms are $2.7$~m in height \newline}
        \begin{tabular}{@{} lccccc @{}} 
        \toprule
        \multicolumn{6}{c}{Simulated training data} \\
        \cmidrule(lr){1-6}
        & Room 1 & Room 2 & Room 3 & Room 4 & Room 5 \\ \cmidrule(lr){2-2} \cmidrule(lr){3-3} \cmidrule(lr){4-4} \cmidrule(lr){5-5} \cmidrule(lr){6-6}
        Room size & ($6\times6$)~m & ($5\times4$)~m & ($10\times6$~m) & ($8\times3$)~m & ($8\times5$)~m\\   
        $\textrm{RT}_{60}$ & $0.3$~s & $0.2$~s & $0.8$~s & $0.4$~s & $0.6$~s\\
        \cmidrule(lr){2-6}
        Signal & \multicolumn{5}{c}{Noiseless signals from WSJ1 \textbf{training} database}\\
        Array position in room & \multicolumn{5}{c}{6 arbitrary positions in each room}\\
        Source-array distance & \multicolumn{5}{c}{$1.5$~m with added noise with $0.1$ variance}\\
        \midrule[\heavyrulewidth]
        \end{tabular}
    
    \label{table:training_data_config}
\end{table}

\begin{table}[t!]
    \centering
    \caption{Configuration of test data generation. All rooms are $3$~m in height \newline}
   
        \begin{tabular}{@{} lcc @{}} 
        \toprule
        \multicolumn{3}{c}{Simulated test data} \\
        \cmidrule(lr){1-3}
        & Room 1 & Room 2\\ 
        \cmidrule(lr){2-2} \cmidrule(lr){3-3}
        Room size & ($5\times7$)~m & ($9\times4$)~m\\   
        $\textrm{RT}_{60}$ & $0.38$~s & $0.7$~s\\
        Source-array distance & $1.3$~m & $1.7$~m\\
        \cmidrule(lr){2-3}
        Signal & \multicolumn{2}{c}{Noiseless signals from WSJ1 \textbf{test} database}\\
        Array position in room & \multicolumn{2}{c}{4 arbitrary positions in each room}\\
        \midrule[\heavyrulewidth]
        \end{tabular}
     \label{table:test_data_config}
\end{table}

For each scenario, two clean signals were randomly drawn from the \ac{WSJ1} database \cite{WSJ} and then convolved with \acp{RIR} corresponding to two different \acp{DOA} in the range $\Theta=\{0,5,\ldots,180\}$. 
The sampling rate of all signals and \acp{RIR} was set to $16\text{KHz}$. The speakers were positioned in a radius of $r=1.5m$ from the center of the microphone array. To enrich the training diversity, the radius of the speakers was perturbed by a Gaussian noise with a variance of  $0.1$~m. The \ac{DOA} of each speaker was calculated \ac{w.r.t.} the center of the microphone array.

 The contributions of the two sources were then summed with a random \ac{SIR} selected in the range of $\text{SIR}\in\left[-2,2\right]$ to obtain the received microphone signals. Next, we calculated the \ac{STFT} of both the mixture and the \ac{STFT} of the separate signals with a frame-length  $K=512$ and an overlap of $75\%$ between two successive frames. 

We then constructed the audio feature matrix ${R}$ as described in Sec.~\ref{subsec:feature_vector}. In the training phase, both the location and a clean recording of each speaker were known, hence they could be used to generate the labels. For each \ac{TF} bin $(l,k)$, the dominating speaker was determined by:
\begin{equation} 
            \mbox{dominant speaker} \leftarrow \argmax_i | s^{i}(l,k)h_{\text{ref}}^{i}(l,k) |.
 \end{equation}
The ground-truth label $D_{l,k}$ is the \ac{DOA} of the dominant speaker. The training set comprised four hours of recordings with 30000 different scenarios of mixtures of two speakers. It is worth noting that as the length of each speaker recording was different, the utterances could  also include non-speech or single-speaker frames. 
The network was trained to minimize the cross-entropy between the correct and the estimated \ac{DOA}. The cross-entropy cost function was summed over all the images in the training set.
The network was implemented in Tensorflow with the ADAM optimizer~\cite{kingma2014adam}.
The number of epochs was set to be 100, and the training stopped after the validation loss increased for 3 successive epochs. The mini-batch size was set to be $64$ images.

\section{Experimental Study}
\label{sec:experimental_study}
\subsection{Experimental setup}

In this section we evaluate the \ac{TF-DOAnet} and compare its performance to classic and DNN-based algorithms.
To objectively evaluate the performance of the \ac{TF-DOAnet}, we first simulated 2 unfamiliar test rooms. Then, we tested our \ac{TF-DOAnet} with real \ac{RIR} recordings in different rooms. Finally, a real-life scenario with fast moving speakers was recorded and tested.

\title{subsec:experimental_setup}
For each test scenario, we selected two speakers from the test set of the \ac{WSJ1} database~\cite{WSJ}, placed them at two different angles between $0^{\circ}$ and $180^{\circ}$ relative to the microphone array, at a distance of either $1$m or $2$m. The signals were generated by convolving the signals with \acp{RIR} corresponding to the source positions and with either simulated or recorded acoustic scenarios.

{\bf Performance measures} Two different measures to objectively evaluate the results were used: the \ac{MAE} and the localization accuracy (Acc.). The \ac{MAE}, computed between the true and estimated \acp{DOA} for each evaluated acoustic condition, is given by 
\begin{equation}
    \label{eq:mae}
    \textrm{MAE}(^\circ)=\frac{1}{N\cdot C}\sum_{c=1}^{C}
      \min_{\pi \in S_N} \sum_{n=1}^{N}|\theta_n^c-\hat{\theta}_{\pi(n)}^c|  ,
\end{equation}
where $N$ is the number of simultaneously active speakers and $C$ is the total number of speech mixture segments considered for evaluation for a specific acoustic condition. In our experiments $N=2$. The true and estimated \acp{DOA} for the
$n$-th speaker in the $c$-th mixture are denoted by $\theta_n^c$ and $\hat{\theta}_n^c$, respectively.

The localization accuracy is given by
\begin{equation}
    \label{eq:accuracy}
        \textrm{Acc.}(\%)=\frac{\hat{C}_{\textrm{acc.}}}{C}\times100
\end{equation}
where $\hat{C}_{\textrm{acc.}}$ denotes the number of speech mixtures for which the localization of the speakers is accurate. We considered the
localization of speakers for a speech frame to be accurate if the distance between the true and the estimated \ac{DOA} for all the speakers was less than or equal to $5^\circ$.

{ \bf Compared algorithms}
We compared the performance of the \ac{TF-DOAnet} with two frequently used baseline methods, namely the \ac{MUSIC} and \ac{SRP-PHAT} algorithms. In addition, we  compared its performance  with the  \ac{CMS-DOA} estimator~\cite{sumitru2019_doaMulti}.\footnote{the trained model is available here \url{https://github.com/Soumitro-Chakrabarty/Single-speaker-localization}}
To facilitate the comparison, the \ac{MUSIC} pseudo-spectrum was computed for each frequency sub-band and for each \ac{STFT} time frame, with an angular resolution of $5^\circ$ over the entire \ac{DOA} domain. Then, it was averaged over all  frequency subbands to obtain a broadband pseudo-spectrum followed by averaging over all the time frames $L$. Next, the two \acp{DOA} with the highest values were selected as the final \ac{DOA} estimates. Similar post-processing was applied to the computed \ac{SRP-PHAT} pseudo-likelihood for each time frame.

\subsection{Speaker localization results}
\paragraph{Static simulated scenario}
We first generated a test dataset with simulated \acp{RIR}. 
Two different rooms were used, as described in Table~\ref{table:test_data_config}. For each scenario, two speakers (male or female) were randomly drawn from the \ac{WSJ1} test database, and placed at two different \acp{DOA} within the range $\{0,5,\ldots,180\}$ relative to the microphone array. The microphone array was similar to the one used in the training phase. Using the \ac{RIR} generator, we generated the \ac{RIR} for the given scenario and convolved it with the speakers' signals. 

The results for the \ac{TF-DOAnet} compared with the competing methods  are depicted in Table~\ref{table:results_simulated}. The tables shows that the deep-learning approaches outperformed the classic approaches. The \ac{TF-DOAnet} achieved very high scores and outperformed the DNN-based \ac{CMS-DOA} algorithm in terms of both \ac{MAE} and accuracy.

\label{subsubsec:static_simulated_scenari}
\paragraph{Static real recordings scenario}
The best way to evaluate the capabilities of the \ac{TF-DOAnet} is testing it with real-life scenarios. For this purpose, we first carried out experiments with real measured \acp{RIR} from a multi-channel impulse response database~\cite{Elior_database}. The database comprises  \acp{RIR} measured in an acoustics lab for three different reverberation times of $\textrm{RT}_{60} = 0.160, 0.360$, and $0.610$~s. The lab dimensions are $6\times 6\times 2.4$~m. 

The recordings were carried out with different \ac{DOA} positions in the range of    $[0^\circ, 180^\circ]$, in steps of $15^\circ$. The sources were positioned at distances of $1$~m and $2$~m from the center of the microphone array.
The recordings were carried out with a linear microphone array consisting of $8$ microphones with three different microphone spacings. For our experiment, we chose
the [8, 8, 8, 8, 8, 8, 8]~cm setup. 
In order to construct an array setup identical to the one in the training phase, we selected a sub-array of the four center microphones out of the total 8 microphones in the original setup. Consequently, we used a \ac{ULA} with $M = 4$ elements with an inter-microphone distance of $8$~cm.

The results for the \ac{TF-DOAnet} compared with the competing methods are depicted in Table~\ref{table:results_measured}. Again, the \ac{TF-DOAnet} outperforms all competing methods, including the \ac{CMS-DOA} algorithm. Interestingly, for the 1~m case, the best results for the \ac{TF-DOAnet} were obtained for the highest reverberation level, namely $\textrm{RT}_{60}=610$~ms, and for the 2~m case, for  $\textrm{RT}_{60}=360$~ms. While surprising at first glance, this can be explained using the following arguments. There is an accumulated evidence that reverberation, if properly addressed, can be beneficial in speech processing, specifically for multi-microphone speech enhancement and source extraction \cite{gannot2001RTF,markovich-golan_multichannel_2009,dokmanic2015raking} and  for speaker localization \cite{deleforge2015acoustic,laufer2016semi}. In reverberant environments, the intricate acoustic propagation pattern constitutes a specific ``fingerprint'' characterizing the location of the speaker(s). When reverberation level increases, this fingerprint becomes more pronounced and is actually more informative than its an-echoic counterpart. An inference methodology that is capable of extracting the essential driving parameters of the \ac{RIR} will therefore improve when the reverberation is higher. If the acoustic propagation becomes even more complex, as is the case of high reverberation and a remote speaker, a slight performance degradation may occur, but as evident from the localization results, for sources located 2~m from the array, the performance for $\textrm{RT}_{60}=610$~ms was still better than the performance for $\textrm{RT}_{60}=160$~ms.

\begin{figure*}[t!]
\centering
	\begin{subfigure}[t]{0.4\textwidth}
	\centering
		\includegraphics[scale=0.225]{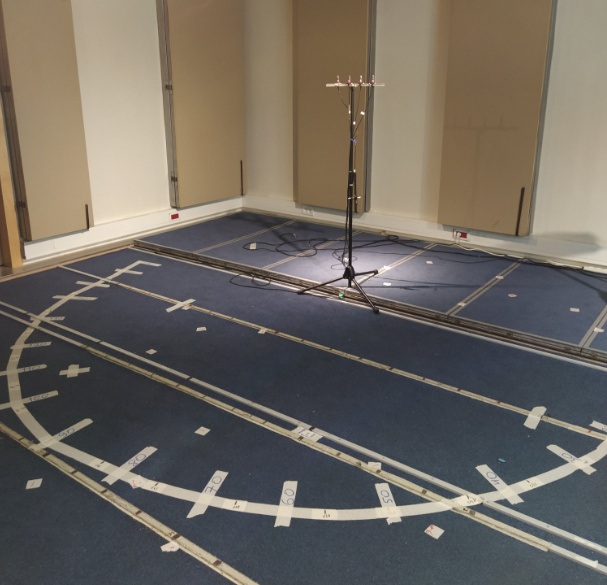}
		\caption{Room view.}
		\label{fig:room}
	\end{subfigure}%
	\begin{subfigure}[t]{0.4\textwidth}
	\centering
		\includegraphics[scale=0.3,trim=0 10 0 0 ,clip,]{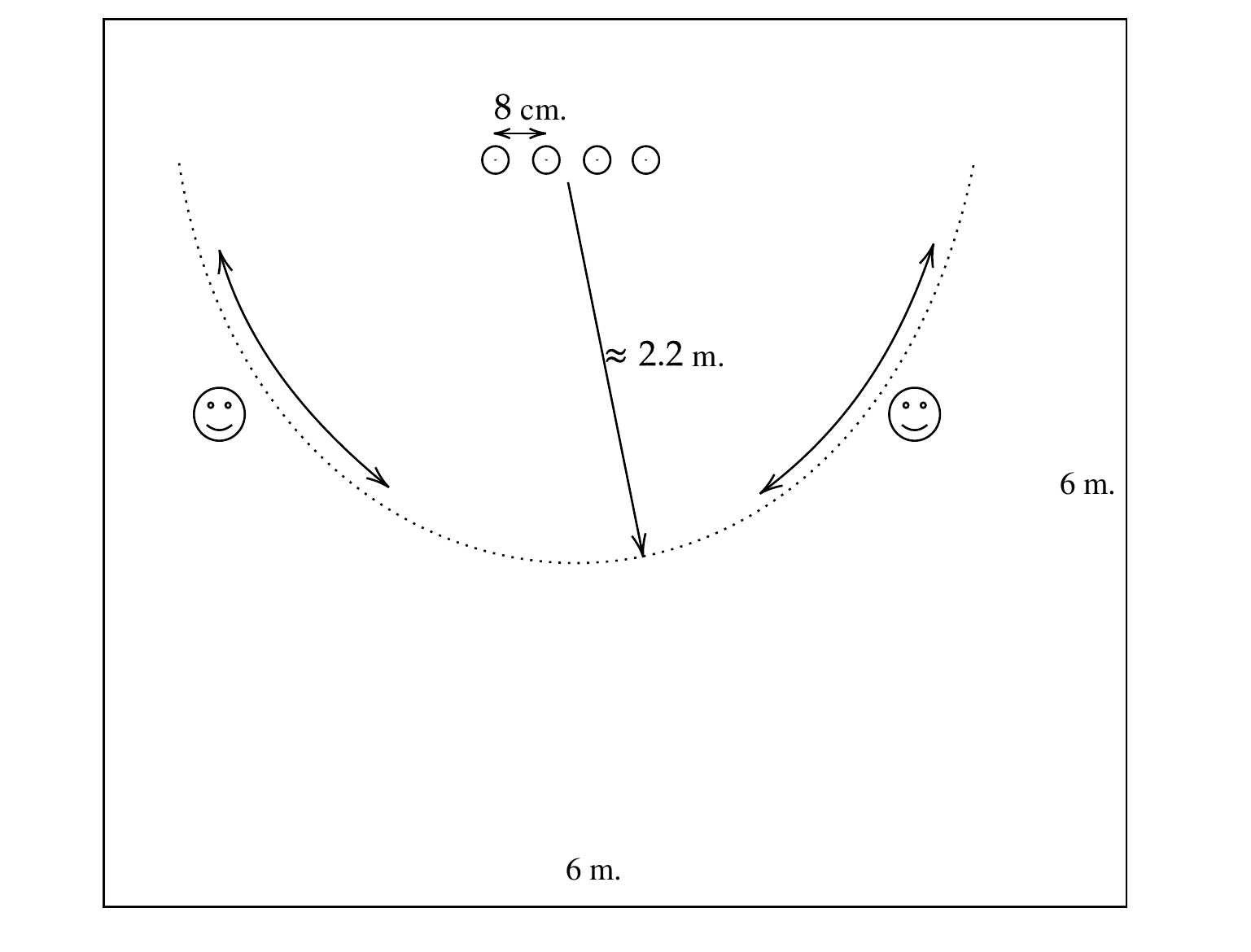}
		\caption{Speakers' trajectory.}
		\label{fig:Room_diagram}
	\end{subfigure}
	\caption{Real-life experiment setup.}
	\label{fig:lab}
\end{figure*}

\paragraph{Real-life dynamic scenario}
To further evaluate the capabilities of the \ac{TF-DOAnet}, we also carried out real dynamic scenarios experiments.
The room dimensions are $6 \times 6 \times 2.4$~m. The room reverberation level can be adjusted and we set the $\textrm{RT}_{60}$ at two levels,  $390$~ms and $720$~ms, respectively. The microphone array consisted of $4$ microphones with an inter-microphone spacing of $8$~cm.  The speakers walked naturally on an arc at a distance of about $2.2$~m from the center of the microphone array. For each $\textrm{RT}_{60}$ two experiments were recorded. The two speakers started at the angles $20^\circ$ and $160^\circ$ and walked until they reached $70^\circ$ and $100^\circ$, respectively, turned around and walked back to their starting point. This was done several times throughout the recording. Figure \ref{fig:room} depicts the real-life experiment setup and Fig.~\ref{fig:Room_diagram} depicts a schematic diagram of the setup of this experiment. The ground truth labels of this experiment were measured with the Marvelmind indoor 3D tracking set.\footnote{\url{https://marvelmind.com/product/starter-set-ia-02-3d/}}

\begin{figure*}[t!]
\centering
	\begin{subfigure}[t]{0.31\textwidth}
	\centering
		\includegraphics[trim=30 0 140 0 ,clip, width=0.9\textwidth]{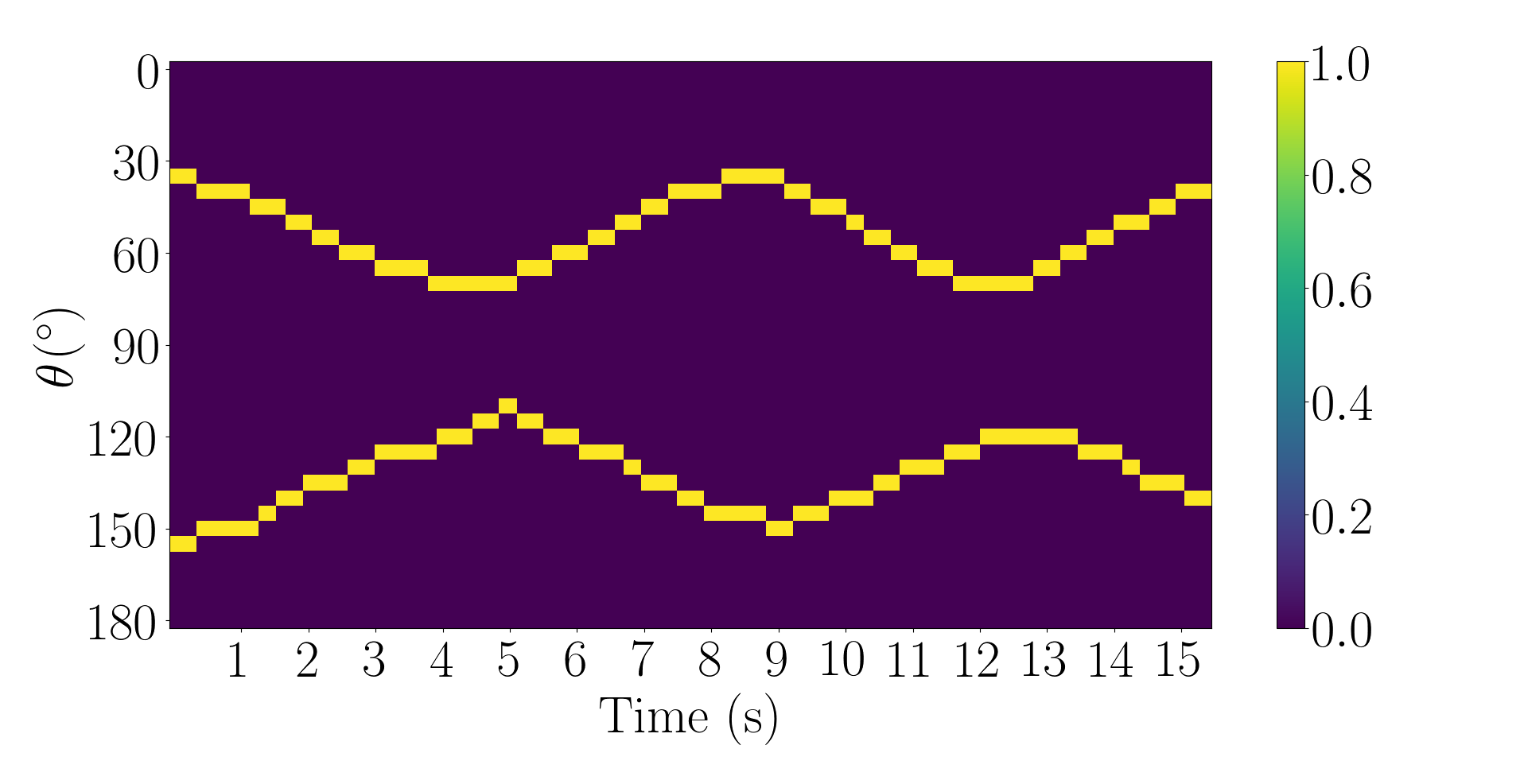}
		\caption{Ground truth.}
		\label{fig:gt}
	\end{subfigure}%
	\begin{subfigure}[t]{0.31\textwidth}
	\centering
		\includegraphics[trim=30 0 140 0 ,clip,width=.9\textwidth]{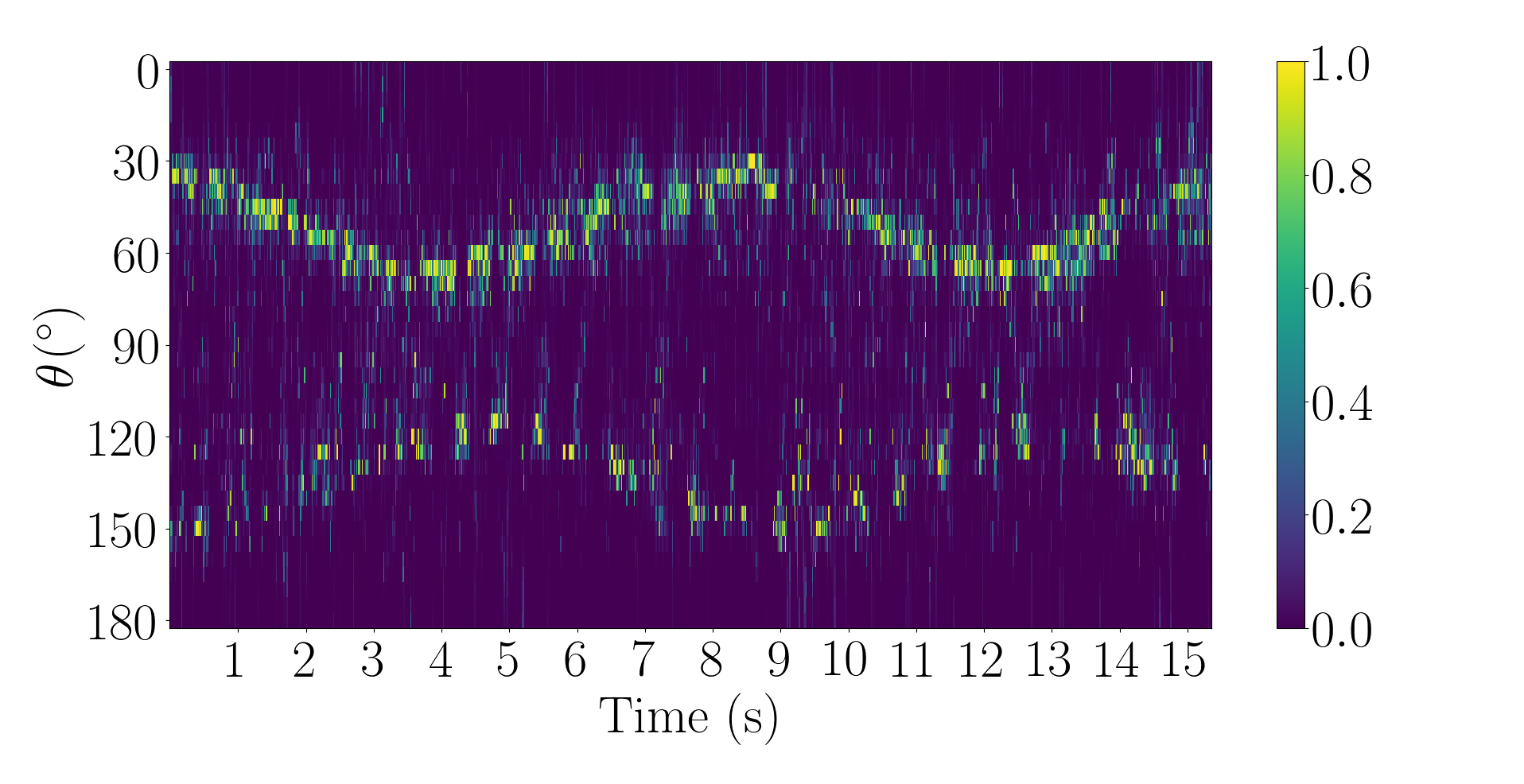}
		\caption{CMS-DOA.}
		\label{fig:somitru}
	\end{subfigure}%
    \begin{subfigure}[t]{0.31\textwidth}
    \centering
		\includegraphics[trim=30 0 140 0 ,clip,width=.9\textwidth]{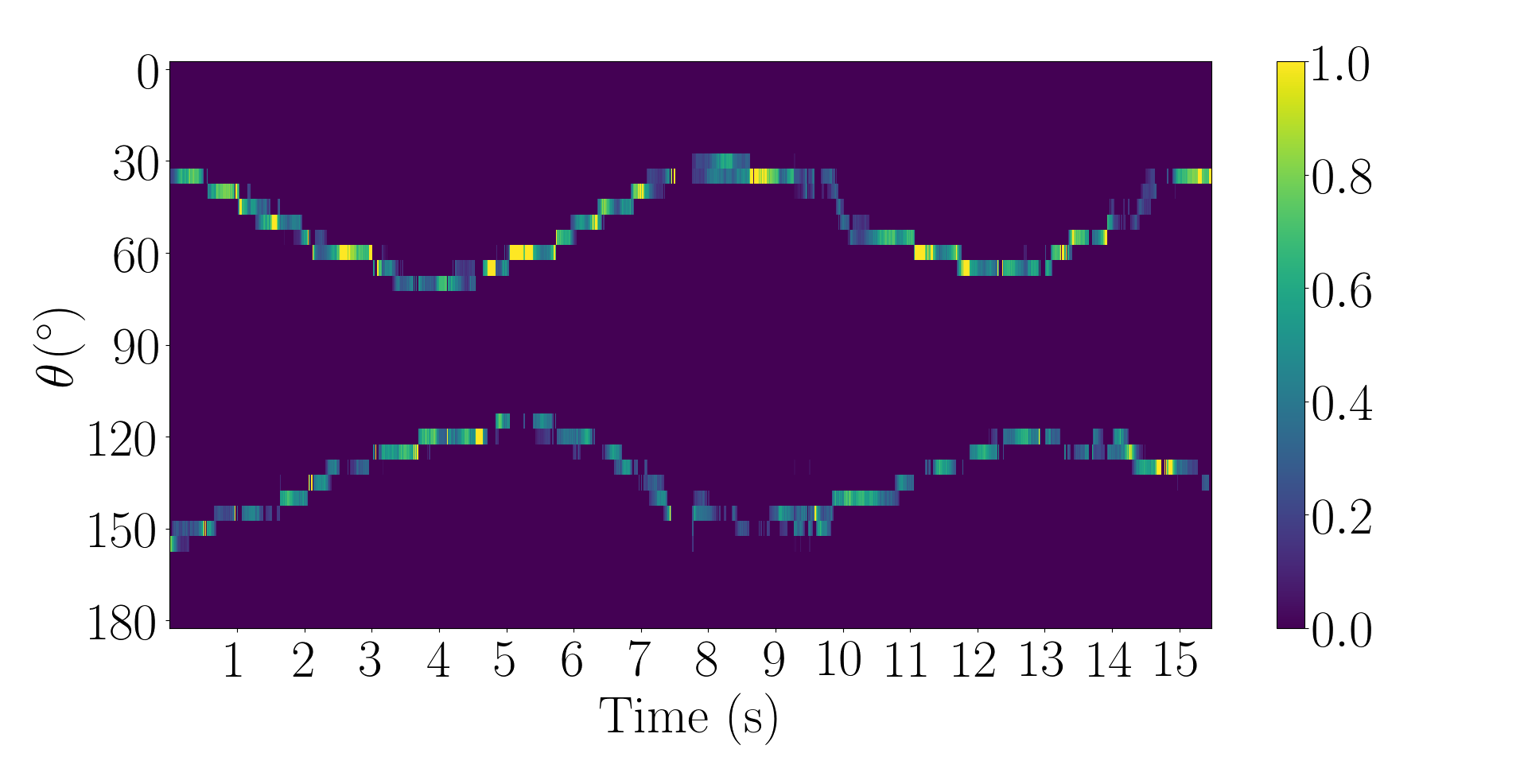}
		\caption{ The \ac{TF-DOAnet}.}
		\label{fig:ours}
	\end{subfigure}
	\caption{Real-life recording of two moving speakers in a $6\times6\times2.4$ room with RT$_{60}=390$ ms.}
	\label{fig:example1}
\centering
	\begin{subfigure}[t]{0.31\textwidth}
	\centering
		\includegraphics[trim=30 0 140 0 ,clip, width=0.9\textwidth]{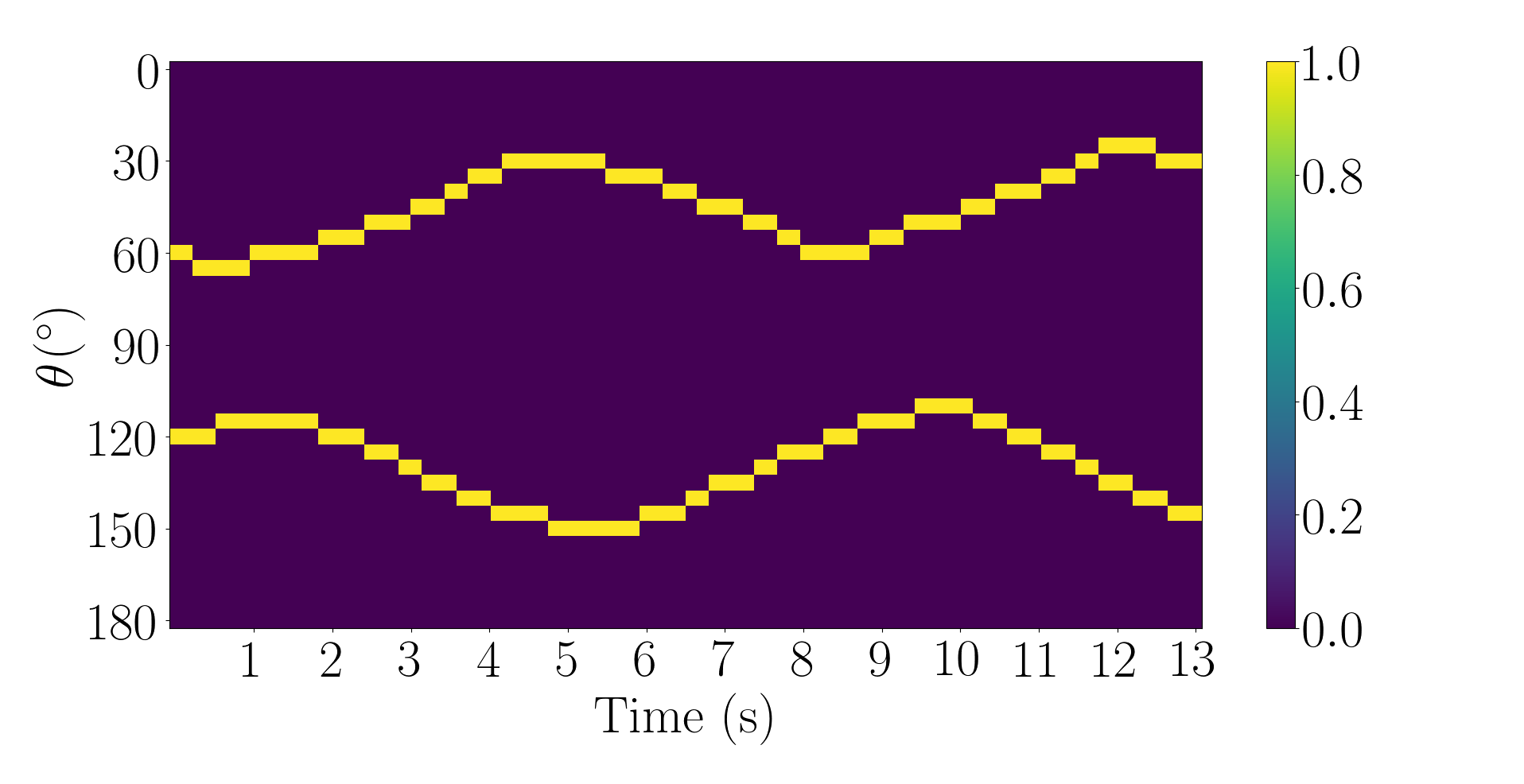}
		\caption{Ground truth.}
		\label{fig:gt}
	\end{subfigure}%
	\begin{subfigure}[t]{0.31\textwidth}
	\centering
		\includegraphics[trim=30 0 140 0 ,clip,width=.9\textwidth]{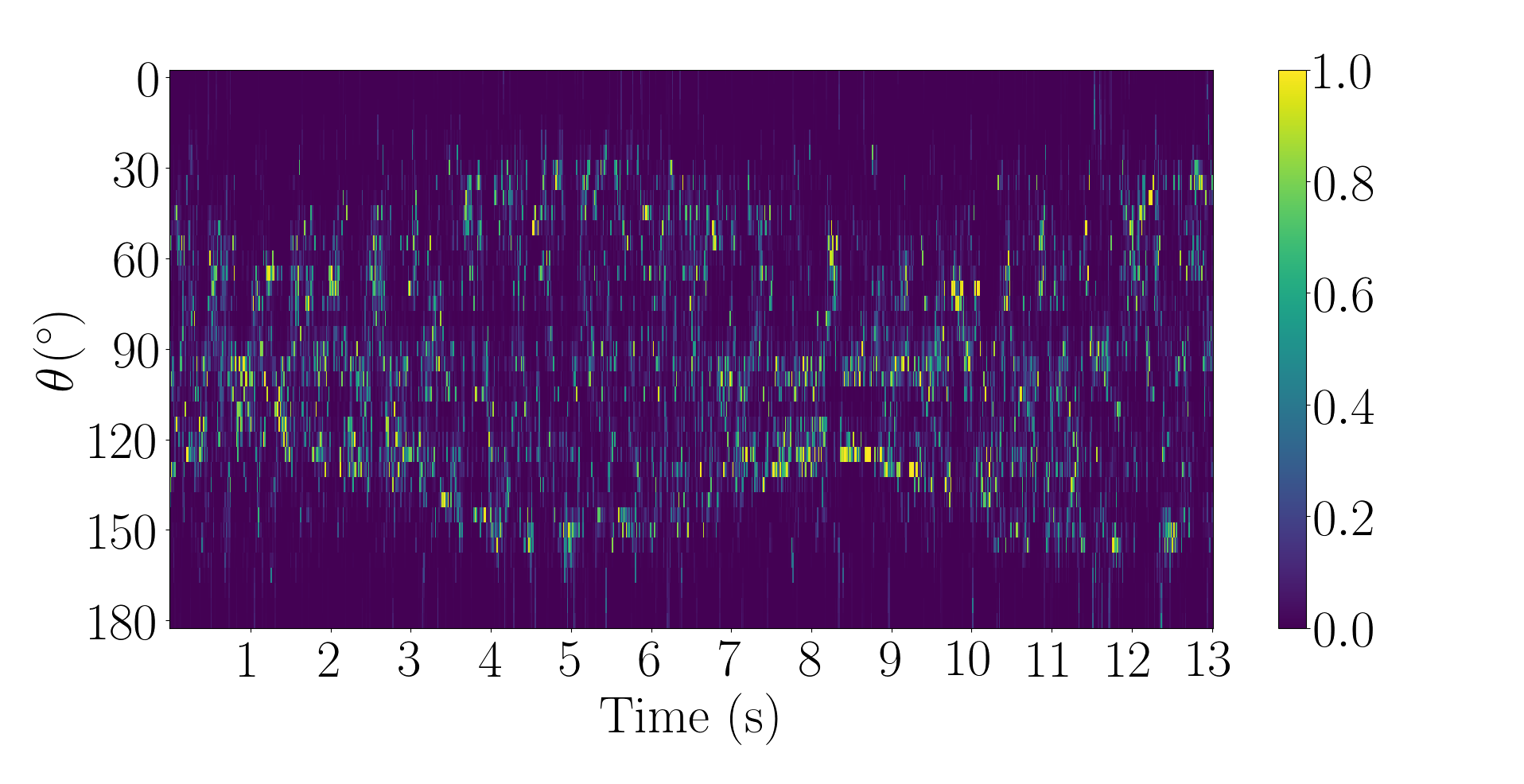}
		\caption{CMS-DOA.}
		\label{fig:somitru}
	\end{subfigure}%
    \begin{subfigure}[t]{0.31\textwidth}
    \centering
		\includegraphics[trim=30 0 140 0 ,clip,width=.9\textwidth]{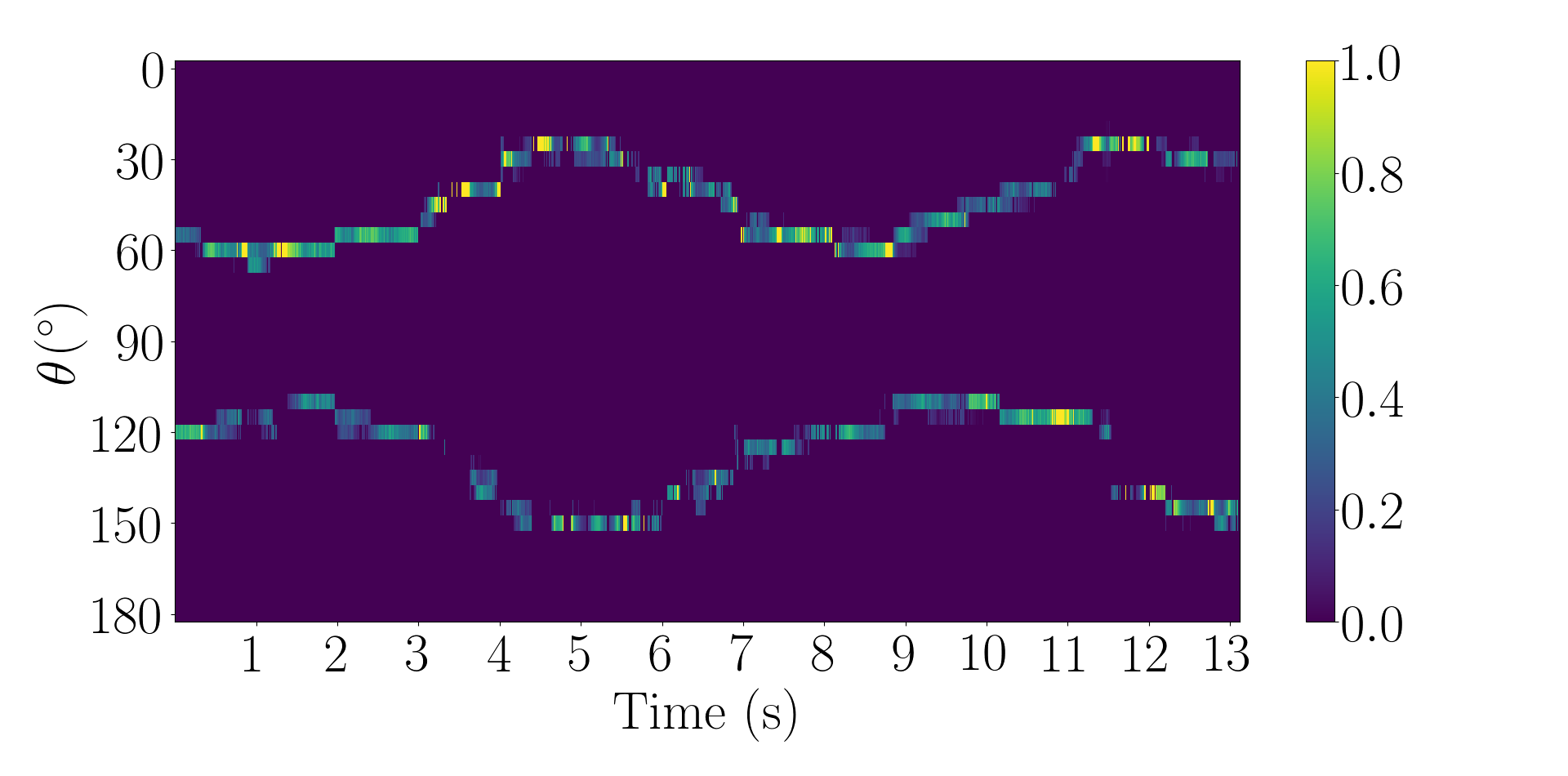}
		\caption{ The \ac{TF-DOAnet}.}
		\label{fig:ours}
	\end{subfigure}
	\caption{Real-life recording of  two moving speakers in a $6\times6\times2.4$ room with RT$_{60}=720$ ms.}
	\label{fig:example2}
\end{figure*}

Figures~\ref{fig:example1} and \ref{fig:example2} depict the results of the two experiments. It is clear that the \ac{TF-DOAnet} outperformed the \ac{CMS-DOA} algorithm, especially for the high RT$_{60}$ conditions. Whereas the \ac{CMS-DOA} fluctuated rapidly, the \ac{TF-DOAnet} output trajectory was smooth and noiseless.

\begin{table}
    \centering
    \caption{Results for two different test rooms with simulated \acp{RIR}\newline}
   
        \begin{tabular}{@{} lcccc @{}} 
        \toprule
        Test Room & \multicolumn{2}{c}{Room 1} & \multicolumn{2}{c}{Room 2}\\ 
        \cmidrule(lr){2-3} \cmidrule(lr){4-5}
        Measure & MAE & Acc. & MAE & Acc.\\
        \cmidrule(lr){2-2} \cmidrule(lr){3-3} \cmidrule(lr){4-4} \cmidrule(lr){5-5}
        MUSIC~\cite{Dmochowski_Benesty_Affes_2007} & 26.2 & 28.4 & 31.5 & 16.9 \\
        SRP-PHAT~\cite{Brandstein_Silverman_1997} & 25.1 & 26.7 & 35.0 & 15.6 \\
        CMS-DOA \cite{sumitru2019_doaMulti} & 13.1 & 71.1 & 24.0 & 38.1 \\
        \ac{TF-DOAnet} & \textbf{0.3} & \textbf{99.5} & \textbf{1.7} & \textbf{94.3} \\
        \midrule[\heavyrulewidth]
        \end{tabular}
     \label{table:results_simulated}
\end{table}

\begin{table}
    \caption{Results for three different rooms at  distances of $1$~m and $2$~m with measured \acp{RIR} \newline}
    \hspace{-1cm}
        \begin{tabular}{@{} lcccccccccccc @{}} 
        \toprule
        Distance & \multicolumn{6}{c}{$1$~m} & \multicolumn{6}{c}{$2$~m}\\ 
        \cmidrule(lr){2-7} \cmidrule(lr){8-13}
        $\textrm{RT}_{60}$ & \multicolumn{2}{c}{0.160~s} & \multicolumn{2}{c}{0.360~s} & \multicolumn{2}{c}{0.610~s} & \multicolumn{2}{c}{0.160~s} & \multicolumn{2}{c}{0.360~s} & \multicolumn{2}{c}{0.610~s} \\
        \cmidrule(lr){2-3} \cmidrule(lr){4-5} \cmidrule(lr){6-7} \cmidrule(lr){8-9} \cmidrule(lr){10-11} \cmidrule(lr){12-13}
        Measure & MAE & Acc. & MAE & Acc. & MAE & Acc. & MAE & Acc. & MAE & Acc. & MAE & Acc. \\
        \cmidrule(lr){2-2} \cmidrule(lr){3-3} \cmidrule(lr){4-4} \cmidrule(lr){5-5} \cmidrule(lr){6-6} \cmidrule(lr){7-7} \cmidrule(lr){8-8} \cmidrule(lr){9-9} \cmidrule(lr){10-10} \cmidrule(lr){11-11} \cmidrule(lr){12-12} \cmidrule(lr){13-13}
        MUSIC & 18.7 & 57.6 & 19.2 & 53.2 &	21.9 & 42.9 & 18.4 & 54.1 & 26.1 & 35.8 & 25.4 & 32.2 \\
        SRP-PHAT & 9.0 & 39.0 & 13.9 & 39.4 & 18.6 & 29.9 & 9.7 & 36.0 & 16.5 & 24.7 & 27.7 & 21.3 \\
        CMS-DOA & 1.6 & 76.3 & 7.3 & 75.2 & 8.4 & 71.9 & 5.1 & 79.5 & 9.7 & 60.1 & 17.5 & 40.0 \\
        \ac{TF-DOAnet} & \textbf{1.3} & \textbf{97.5} & \textbf{3.5} & \textbf{83.5} & \textbf{0.9} & \textbf{98.3} & \textbf{5.0} & \textbf{89.5} & \textbf{1.7} & \textbf{95.7} & \textbf{4.8} & \textbf{84.2} \\
        \midrule[\heavyrulewidth]
        \end{tabular}
    
    \label{table:results_measured}
\end{table} 

\subsection{Ablation study}
In our implementation, we used the real and imaginary part of the \ac{RTF}~\eqref{eq:est_RTF}. Other approaches might be beneficial. For example, in \cite{chazan_hammer_hazan_goldberger_gannot_2019}, the $\cos$ and the $\sin$ of the phase of the \ac{RTF} were used. In other approaches, the spectrum was added to the spatial features~\cite{hershi_multi_channel_deep_cluster}.  

In this section, the different features were tested with the same model. We compared the proposed features with two other features. First, we used the proposed features as described in~\eqref{eq:est_RTF}. The second approach was a variant of our approach with the  spectrum added (`TF-DOAnet with Spec.'). The third, used the $\cos$ and the $\sin$ features as presented in \cite{chazan_hammer_hazan_goldberger_gannot_2019} (`Cos-Sin').
All features were crafted from the same training data described in Sec.~\ref{subsec:training_phase}. We tested the different approaches in the test conditions described in~\ref{table:test_data_config}. 

First, it is clear that all the features with our high resolution \ac{TF} model outperformed the frame-based \ac{CMS-DOA} algorithm, as reported in Table \ref{table:results_simulated}. This confirms that the \ac{TF} supervision is beneficial for the task at hand. Second, the proposed features were shown to be better than the Cos-Sin features. Finally, it is very interesting to note that the addition of the spectrum features slightly deteriorated the results for this task.  

\begin{table}[!htbp]
    \centering
    \caption{Ablation study results with different features \newline}
    
        \begin{tabular}{@{} lcccc @{}} 
        \toprule
        Test Room & \multicolumn{2}{c}{Room 1} & \multicolumn{2}{c}{Room 2}\\ 
        \cmidrule(lr){2-3} \cmidrule(lr){4-5}
        Measure & MAE & Acc. & MAE & Acc.\\
        \cmidrule(lr){2-2} \cmidrule(lr){3-3} \cmidrule(lr){4-4} \cmidrule(lr){5-5}
        Cos-Sin & 1.2 & 96.1 & 2.8 & 91.3 \\
        \ac{TF-DOAnet}  with Spec. & 0.6 & 98.4 & 3.3 & 86.7 \\
        \ac{TF-DOAnet}  & \textbf{0.3} & \textbf{99.5} & \textbf{1.7} & \textbf{94.3} \\
        \midrule[\heavyrulewidth]
        \end{tabular}
    \label{table:ablation}
\end{table}

\section{Conclusions}
A \ac{FCN} approach was presented in this paper for the \ac{DOA} estimation task. Instantaneous \ac{RTF} features were used to train the model. The high \ac{TF} resolution facilitated the tracking of multiple moving speakers simultaneously. A comprehensive experimental study was carried out with simulated and real-life recordings. The proposed approach outperformed both the classic and \ac{CNN}-based \ac{SOTA} algorithms in all experiments.
Training and test datasets which represent different real-life scenarios were constructed as a \ac{DOA} benchmark and will become available after publication.

\section*{Broader impact}
Several modern technologies can benefit from the proposed localization algorithm. We already mentioned the emerging technology of smart speakers in the Introduction. These devices are equipped with multiple microphones and are implementing location-specific tasks, e.g.~the extraction of the speaker of interest.
Of particular interest are socially assistive robots (SARs), as they are likely to play an important role in healthcare and psychological well-being, in particular during non-medical phases inherent to any hospital process. 

The algorithm neither uses the content nor the identity of the speakers and hence does not to violate the privacy of the users. Moreover, since normally speech signal cannot propagate over long distances, the algorithm application is limited to small enclosures. 
\bibliographystyle{plain}
\bibliography{main}

\end{document}